\documentclass[aps,pra,twocolumn,showpacs,preprintnumbers,amsmath,superscriptaddress,amssymb,floats,nofootinbib]{revtex4-1}

\usepackage{graphicx}
\usepackage{dcolumn}
\usepackage{subfigure}

\def\PRL{\rm Phys. Rev. Lett.}

\def\PRD{{\rm Phys. Rev.} D}
\def\PRA{{\rm Phys. Rev.} A}

\def\PR{\rm Phys. Rev.}

\def\MRM{\rm Magn. Res. Med.}

\def\LTP{\rm J. Low Temp. Phys.}

\def\MRM{\rm Mag. Res. Med.}
\def\JAP{\rm J. Appl. Phys.}
\def\Journal#1#2#3#4{{#1} {\bf #2}, #3 (#4)}

\begin{document}

\title{\large \bf Pressure Dependent Wall Relaxation in Polarized $^3$He Gaseous Cells}
\author{C. Peng, W. Zheng, P. -H. Chu, H. Gao, Y. Zhang \\
\textit{Triangle Universities Nuclear Laboratory and Department of Physics, Duke University, Durham, NC 27708, USA}}
\noaffiliation

\begin{abstract}
	Pressure dependence of longitudinal relaxation time (T$_1$) due to the cell wall was observed previously at both room temperature and low temperature in valved Rb-coated refillable $^3$He gaseous cells in \cite{Zheng2}. The diffusion of $^3$He from measurement cell through a capillary tube to the valve and the subsequent depolarization on the surface of the valve was proposed to possibly explain such a pressure dependence at room temperature \cite{Saam}. In this paper, we investigate this diffusion effect through measurements of T$_1$ with newly designed Rb-coated Pyrex glass cells at 295 K as well as finite element analysis (FEA) studies. Both the experimental results and FEA studies show that the diffusion effect is insufficient to explain the observed linear pressure-dependent behavior of T$_1$.
\end{abstract}

\maketitle

Spin polarized $^3$He gas has been used widely in lepton scattering experiments as an effective polarized neutron target \cite{Xu}, as a signal source used in Magnetic Resonance Imaging for diagnostics of lung airways \cite{Middleton2}, and more recently in searches of exotic spin-dependent interactions \cite{Petukhov, Zheng, Chu}. A quantity important to the production and storage of polarized $^3$He for such applications is the spin lattice relaxation time T$_1$, also known as the longitudinal relaxation time. Several factors contribute to this longitudinal relaxation, and the most significant ones are $^3$He dipole-dipole interaction \cite{Newbury,Mullin}, magnetic field gradient induced relaxation \cite{Cates}, and wall relaxation. Among these effects, the relaxation due to the wall is the least understood and the most difficult to control.

A widely accepted model for relaxation of polarized gas inside a cell due to paramagnetic sites in the cell wall is expressed as \cite{Garwin}
\begin{equation}
	\frac{1}{T_1}=\frac{\frac{1}{4}\int\mu n \bar{v}dS}{\int n dV}=\frac{\mu\bar{v}S}{4V},
	\label{eq:oldT1}
\end{equation}
where $S$ and $V$ represent the total surface and volume of the cell, respectively; $\bar{v}$ is the gas mean velocity, $n$ is the number density of the gas, and $\mu$ is the depolarization probability per collision due to paramagnetic impurities. This model shows that as long as gas density or pressure has no effects on $\mu$, T$_1$ does not depend upon the pressure of the gas.

Recently, Zheng \textit{et al.} \cite{Zheng2} observed a linear pressure dependence of T$_1$ at 4.2 K and 295 K in gaseous $^3$He cells made of bare Pyrex glass or Cs- or Rb-coated Pyrex. They interpreted that, in contrary to Eq. (\ref{eq:oldT1}), T$_1$ is in general pressure dependent from paramagnetic wall relaxation in the cell. Saam \textit{et al.} \cite{Saam} disagreed with this interpretation, and proposed that depolarization on the valve and capillary tube connecting the cell to the valve could significantly contribute to the relaxation. Hence, the observed pressure dependent relaxation at room temperature was explained in \cite{Saam} as diffusion of $^3$He through the capillary tube to the valve and subsequent depolarization on the surface of the valve.

In this paper, we report results from measurements using newly designed Rb-coated Pyrex glass cells at 295 K, and results from finite element analysis (FEA) studies of these measurements and the experiment reported in \cite{Zheng2}. The results from both the new experiment and FEA studies do not support the interpretation proposed in \cite{Saam}, showing that the observation of the pressure dependent T$_1$ remains an open question.

The boundary condition in \cite{Saam} describes the surface paramagnetic relaxation in the following way: the macroscopic magnetic flux due to diffusion at the wall is set to equal to the number of collisions per second times the depolarization probability $\mu$,
\begin{equation}
	D\frac{\partial\rho(r,t)}{\partial r}|_{r=R}=-\frac{\bar{v}}{4}\mu\rho(r,t),
	\label{eq:bc}
\end{equation}
where $D$ is the diffusion coefficient, $R$ is the radius of the cell, and $\rho(r,t)$ is the density of polarized $^3$He. Under this boundary condition, when some surface is sufficiently dirty so that $^3$He spins are depolarized once they hit the wall, the diffusion time becomes the dominant factor contributing to T$_1$, and thus results in the pressure dependence of the paramagnetic wall relaxation. Based on the derivation in \cite{Saam}, the pressure dependence observed in the room temperature T$_1$ measurement in \cite{Zheng2} could be solely ascribed to the diffusion to the o-ring valve used in the Rb-coated cells. Assuming the valve surface has a very large depolarization probability, the depolarization of $^3$He spins at other parts of the cell can be ignored. Based on this assumption, i.e. $\mu=1$ on the valve, the relaxation time due to the diffusion effect is denoted as $(T_1)_{val}$ and given by a simplified formula in \cite{Jacob}:
\begin{equation}
	\left(\frac{1}{T_1}\right)_{val}=\frac{\pi r^2 D}{V L},
    \label{eq:tc}
\end{equation}
where $V$ is the cell volume, $r$ and $L$ represent the inner radius and the length of the capillary tube, respectively. The estimation of $(T_1)_{val}$ using this formula is 1.2 h for $D=12$ cm$^2$/s, close to the measured T$_1$ of 2 h at 0.15 atm in \cite{Zheng2}.

\begin{figure}
	\centering
	\includegraphics[width=0.5\textwidth]{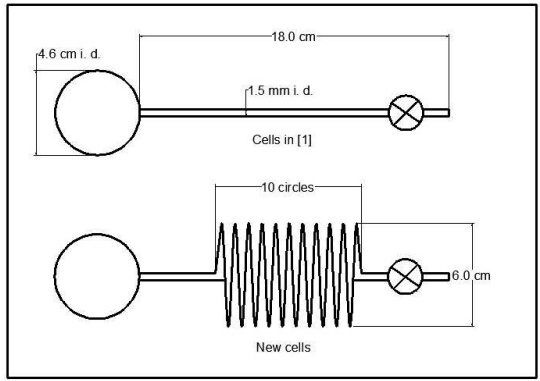}
	\caption{The designs of the Rb-coated cells used in \cite{Zheng2} (top) and in current experiment (bottom). The new cell has a spiral capillary tube to the refilling valve. The spiral tube has 10 turns in total, and the outer diameter of each turn is 6 cm, resulting in a total length of approximately 195 cm.}
	\label{fig:Newcells}
\end{figure}

To further investigate the interpretation in \cite{Saam}, we have carried out an experiment employing two new cells with lengthened capillary tubes. As shown in Eq. (\ref{eq:tc}), the diffusion effect can be reduced by either lengthening or narrowing the capillary tube. Since Rb at various stages of the experiment may block a further narrowed capillary, the new cells were designed to have a significantly lengthened capillary tube. Hence, the diffusion effect through the capillary tube to the refilling valve would be minimized to suppress the relaxation due to depolarization on the valve surface. Fig. \ref{fig:Newcells} shows the schematic of the new cells in comparison with cells in \cite{Zheng2}. The cells are made of Pyrex glass; its spherical volume is Rb-coated. Each of the new cells has a spiral capillary tube connecting the spherical volume to the o-ring refilling valve. The inner diameter of the spherical volume and the capillary tube are 46 mm and 1.5 mm, respectively. The total length of the capillary tube is approximately 195 cm, while the length of the capillary tube of the cells used for the room temperature measurements in \cite{Zheng2} is 18 cm.

T$_1$ for new cells were measured at 295 K utilizing the free induction decay (FID) technique at 22.5 kHz. These detachable cells were filled with $^3$He to 1 atm or 1.5 atm via a $^3$He/N$_2$ gas handling system, and then diluted to 0.2 atm, 0.3 atm or 0.45 atm using different dilution volumes \cite{Zheng2}. For each value of pressure, T$_1$ was measured between 2 and 5 times; the average value was used as the measured T$_1$, and the standard deviation for the measurements at the same pressure was less than 5\%. The magnetic holding field was provided by a Helmholtz coil pair with a field gradient of $<$ 2.3 mG/cm. The gradient induced T$_1$ was more than one thousand hours and thus was negligible. To extract T$_1$ due to wall relaxation from the measured value, the dipole-dipole induced T$_1$ \cite{Newbury,Mullin} was calculated and subtracted. Therefore, the discussion below pertains only to the T$_1$ due to wall relaxation.

\begin{figure}
\centering
\subfigure{
   \includegraphics[width=0.44\textwidth]{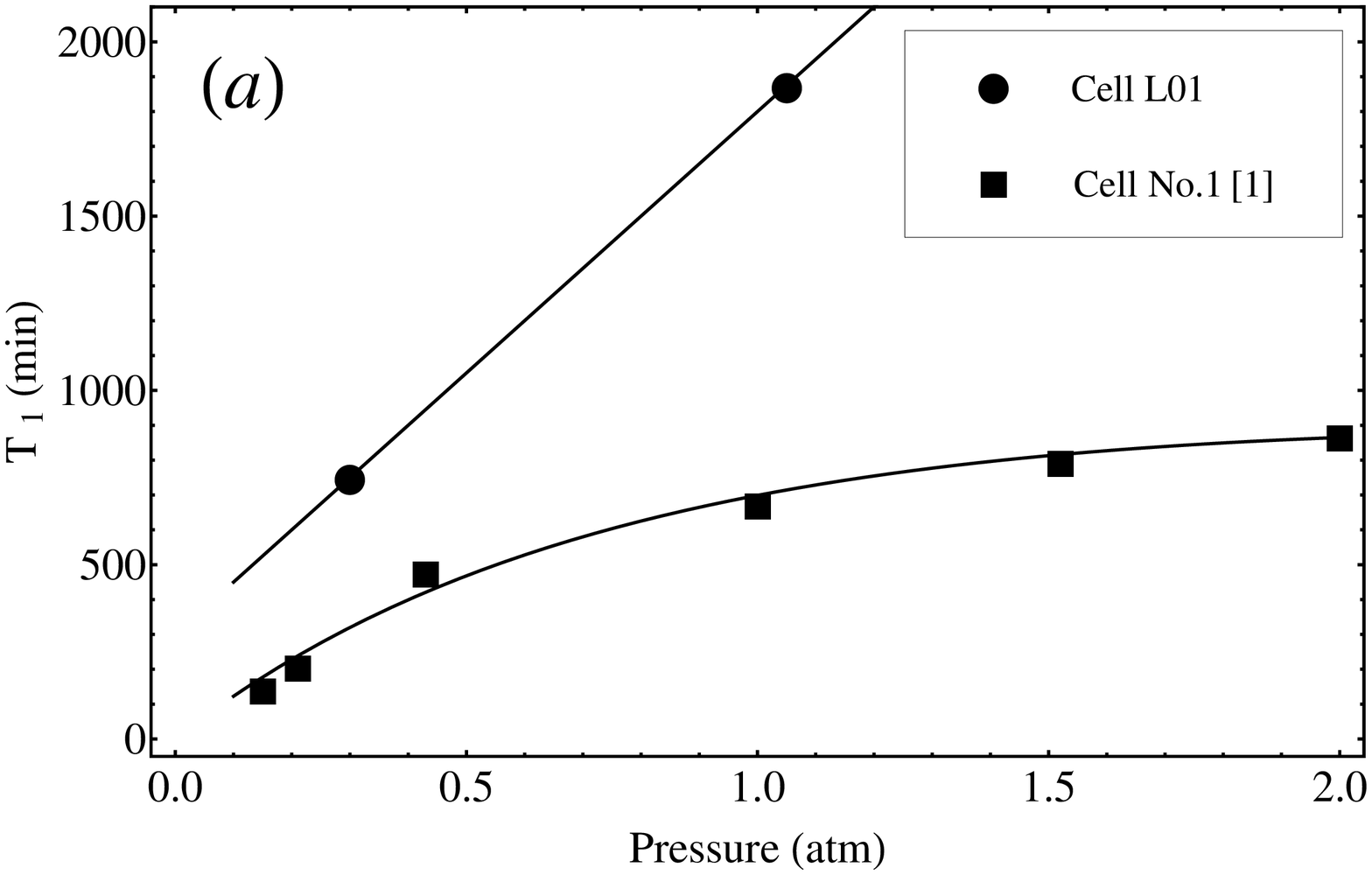}
   \label{fig:T1plots_1} }
\subfigure{
   \includegraphics[width=0.44\textwidth]{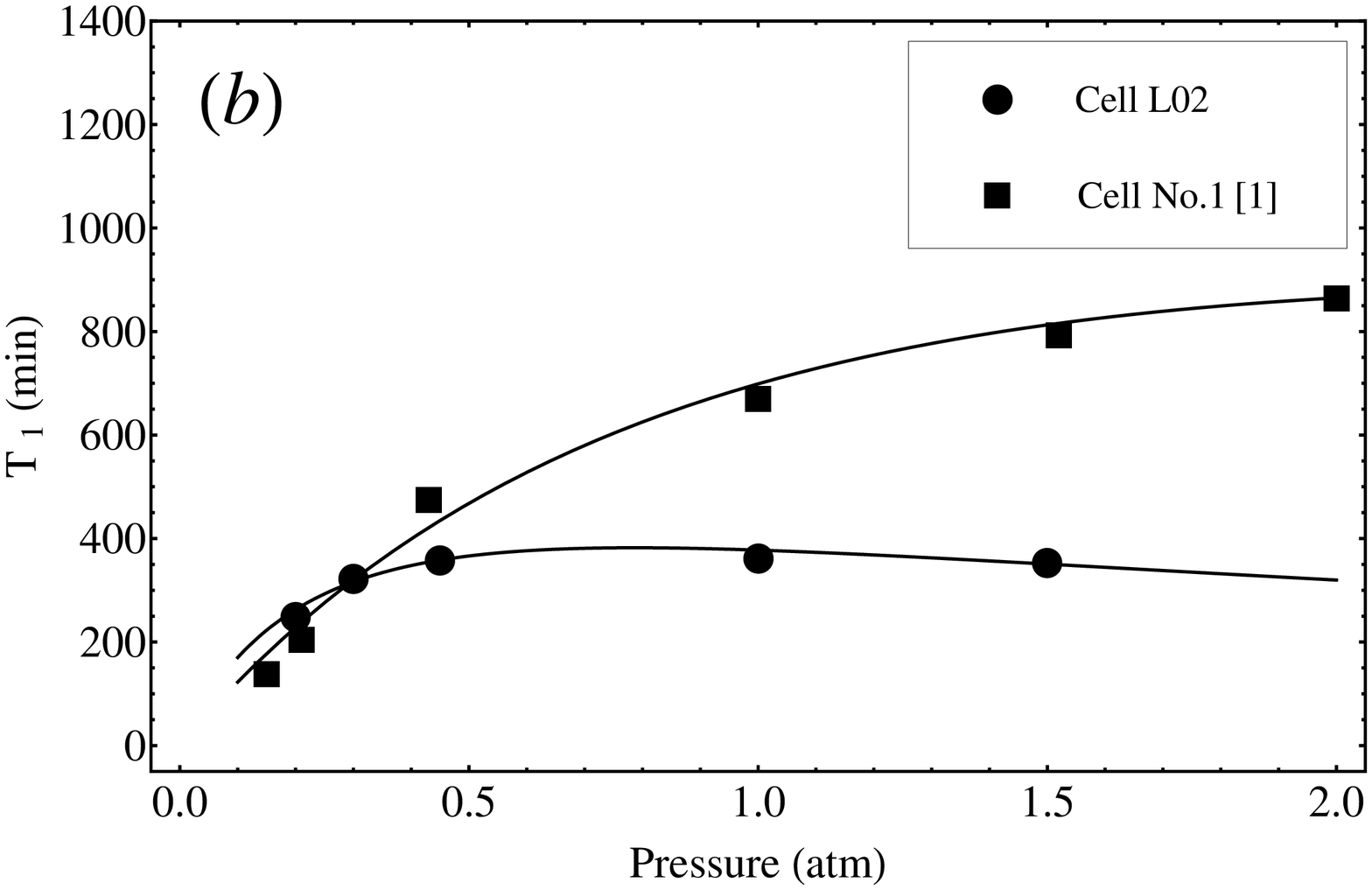}
   \label{fig:T1plots_2} }
\caption{T$_1$ of three different Rb-coated cells at room temperature. (a) shows the T$_1$ of cell L01 and No.1, and (b) shows the T1 of cell L02 and No.1. The data of L01 is fitted linearly, while the other data sets are fitted by the Eq. (\ref{eq:ff}) in \cite{Zheng2}.}
\label{fig:T1plots}
\end{figure}

Fig. \ref{fig:T1plots} shows the T$_1$ of the new cells combined with the experimental data of cell No.1 at 295 K in \cite{Zheng2}. The two new cells with lengthened capillary tubes are named L01 and L02. L01 has a long T$_1$ on the order of tens of hours, while the T$_1$ for L02 ranges from 4 to 6 hours. This difference in T$_1$ between these two cells may be caused by different surface conditions or unexpected contaminations in the glass-cleaning process. Despite this difference, a pressure-dependent behavior in T$_1$ was observed in both cells, as shown in Fig. \ref{fig:T1plots}. The T$_1$ of L01 is 1860 minutes at 1.0 atm, and then decreases to 750 minutes when the $^3$He pressure is diluted to 0.3 atm. The cell was unfortunately contaminated due to an accident in the refilling procedure after these two measurements, but the data already indicate a positive correlation between T$_1$ and $^3$He pressure. The T$_1$ of L02 was measured at 5 different values of pressure, and the results were fit to the following expression in \cite{Zheng2}:
\begin{equation}
	\frac{1}{T_1}=\frac{1}{c_1 p} + \frac{1}{c_2}+\frac{p}{c_3},
	\label{eq:ff}
\end{equation}
where $c_1$, $c_2$, and $c_3$ are fitting parameters, and $p$ is the pressure of $^3$He. The $T_1$ behavior of cell L02 is well described by Eq. \ref{eq:ff} as shown in Fig. \ref{fig:T1plots}(b), and it is similar to that of cell No.1 in \cite{Zheng2}. Both cells L02 and No.1, show a T$_1$ that increases with the rise of the $^3$He pressure when the pressure is low, and then flattens out when the pressure increases further. The results from these two new cells suggest again a linearly pressure-dependent relaxation behavior at low pressure as observed in \cite{Zheng2} .

According to the interpretation proposed in \cite{Saam}, the pressure dependent relaxation observed here can be ascribed to the diffusion effect through the capillary tube to the valve. Using the parameters in \cite{Saam} together with the new dimensions of the cells (L = 195 cm), the estimate of (T$_1$)$_{val}$ is approximately 26 hours at 0.3 atm for cell L01 and L02. Given the fact that the measured T$_1$ of L01 is 12.5 hours at 0.3 atm, this estimated value (T$_1$)$_{val}$ seems to be a reasonable candidate for the pressure dependent relaxation of cell L01. However, further analysis described below about the behavior of all values of T$_1$ shown in Fig. \ref{fig:T1plots} demonstrates that (T$_1$)$_{val}$ can not be the dominant factor determining the pressure dependent relaxation. Combining Eq. (\ref{eq:tc}) and (\ref{eq:ff}), we assume the linear pressure-dependent term of T$_1$ satisfies
\begin{equation}
	\frac{1}{c_1p}=\frac{1}{c_1'p}+\frac{1}{CL},
    \label{eq:c1}
\end{equation}
where $C$ is equal to $V/(\pi r^2 D)$, and is proportional to the pressure through $1/D$; $L$ is the length of the capillary tube. $1/CL$ term represents the contribution from (T$_1$)$_{val}$, and $1/c_1'$ represents the contributions from other possible sources of the pressure dependent relaxation. If (T$_1$)$_{val}$ is the dominant factor, i.e., $1/c_1p \approx 1/CL$, c$_1$ will be approximately proportional to the length of the capillary tube. However, the fitted values of c$_1$ of all the cells shown in Fig. \ref{fig:T1plots} are close to each other, despite the fact that the length of the capillary tube is increased by more than 10 times for cell L01 and L02. Fitting with Eq. (\ref{eq:ff}), we obtain that c$_1$= 1288 $\pm$ 110 for cell No.1; and that c$_1$ = 1611 and 2335 $\pm$ 484 min/atm for L01 and L02, respectively \footnote{L01 is a linear fit.}. Our results show that the c$_1$ is not solely determined by the length of the capillary tube, and the diffusion effect thus is not dominant in the observed pressure dependent relaxation.

\begin{figure}
    \centering
	\includegraphics[width=0.44\textwidth]{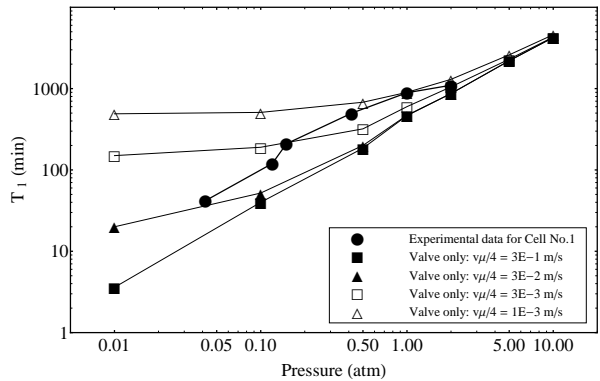}
	\caption{FEA study using the geometry of the Rb-coated cell reported in \cite{Zheng2}. This study assumes that the depolarization only takes place at the valve. Results with four different depolarization probabilities are plotted.}
	\label{fig:T1vsP}
\end{figure}

In this work, we have carried out finite element analysis (FEA) studies to investigate whether the interpretation proposed in \cite{Saam} can describe the experimental results for the cell reported in \cite{Zheng2} and the new cells. In the FEA study of the Rb-coated cell in \cite{Zheng2}, we use the exact geometry of the cell and assume that the depolarization only happens on the valve. Using the boundary condition shown in Eq. (\ref{eq:bc}), we have calculated T$_1$ for four different values of $\bar{v}\mu/4$ from $1\times 10^{-3}$ m/s to $3\times 10^{-1}$ m/s. Since the mean velocity of $^3$He atoms at room temperature is about 1500 m/s, this corresponds to the depolarization probability $\mu$ from $2.7\times 10^{-6}$ to $8\times 10^{-4}$. The results of this FEA study and the experimental data are shown in Fig. \ref{fig:T1vsP}.

\begin{figure}
\centering
\subfigure{
   \includegraphics[width=0.44\textwidth]{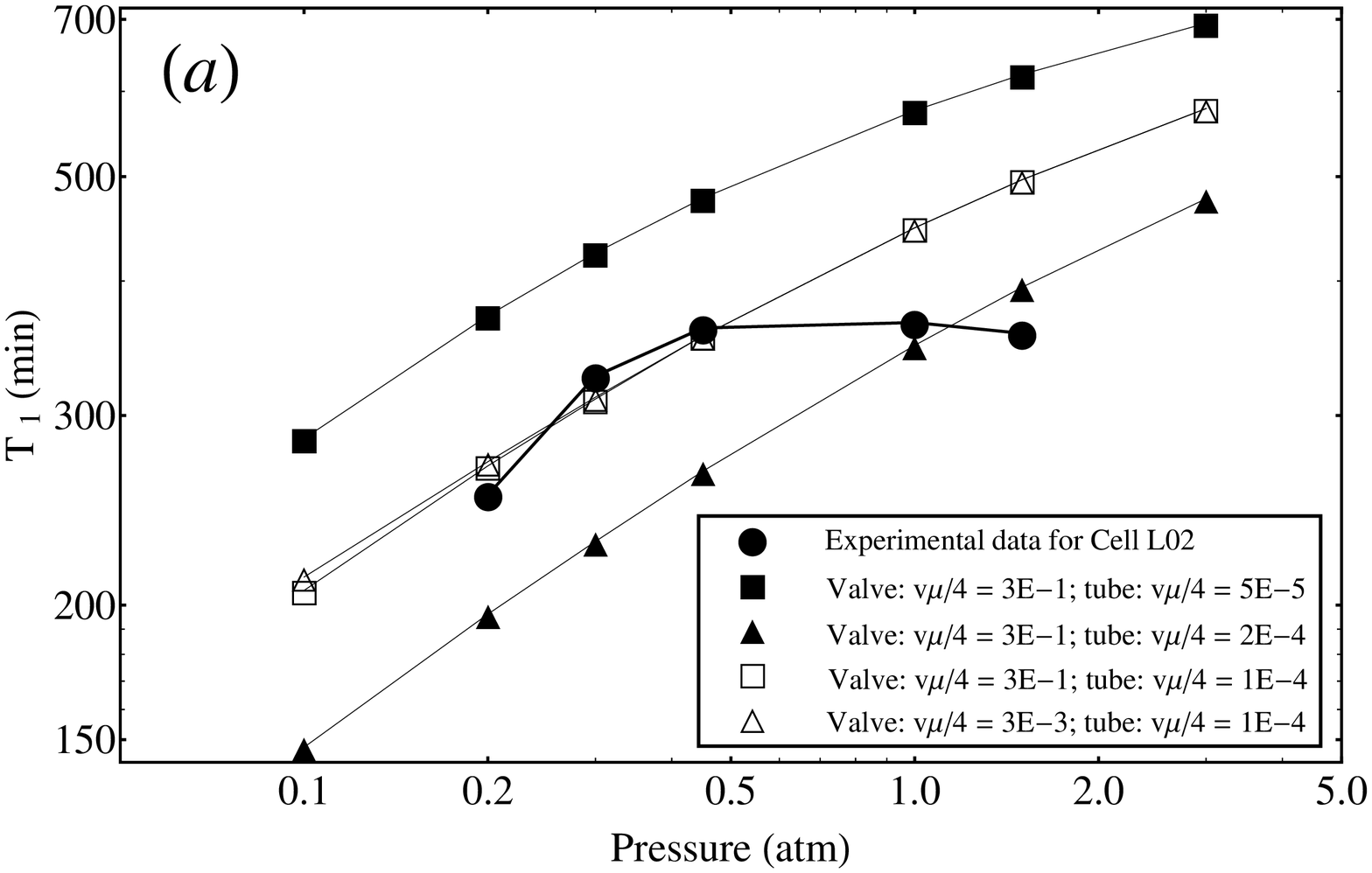}
   \label{fig:T1plots_1} }
\subfigure{
   \includegraphics[width=0.44\textwidth]{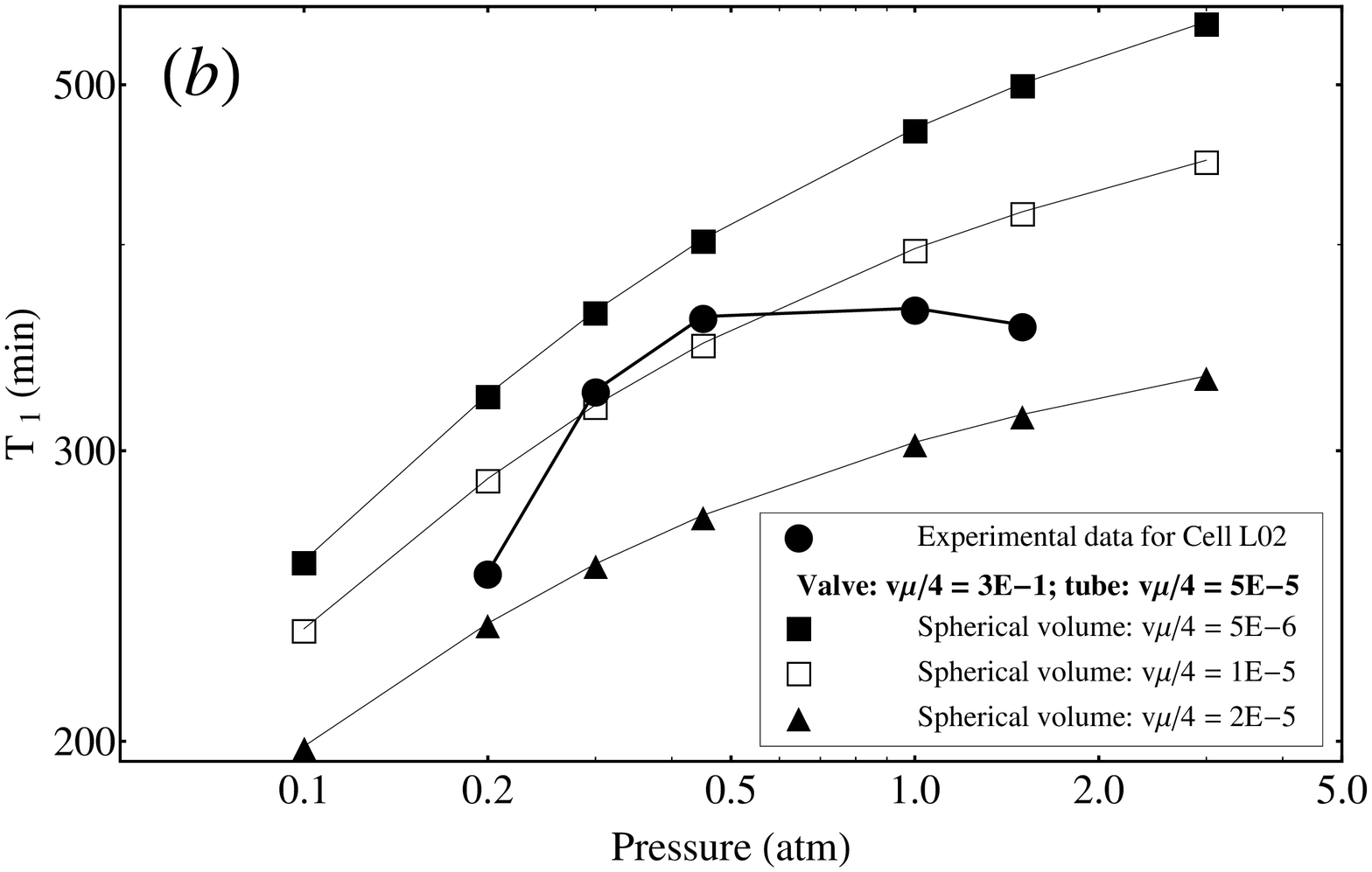}
   \label{fig:T1plots_2} }
	\caption{The FEA calculations of T$_1$ for cell L02. This study assumes that the depolarization takes place on the surfaces of the valve and the capillary tube. (a) shows the calculations assuming depolarization on the surface of the spherical volume surface can be neglected, while (b) shows the calculations including the depolarization from the surface of the spherical volume.}
	\label{fig:NewFEA}
\end{figure}

Contrary to experimental data, the calculation using the boundary condition of Eq. (\ref{eq:bc}) shows a linear pressure dependence at high pressure regardless of the magnitude of the depolarization probability. When pressure is low, T$_1$ with small depolarization probabilities tends to be a constant independent of the pressure. This is opposite to the observation in the experiment, where the linear pressure dependence is observed at low pressures. With a larger depolarization probability on the valve, the linear pressure dependence tends to extend more to the low pressure region. However, the predicted T$_1$ in this case will be much smaller compared to the experimental values.

In the case of the new cells, the situation can be more complicated than the assumption that the depolarization takes place only at the valve. For example, the surface condition of the capillary tube is quite likely inferior to that of the spherical volume, where in the latter case it is improved by rubidium coating. This possible source of relaxation becomes significant for the new cells with the lengthened capillary tube, because its surface area of the tube is comparable to that of the spherical volume. To evaluate the relaxation on the tube surface, a FEA study using the boundary condition of Eq. (\ref{eq:bc}) for cell L02 is carried out. As shown in Fig. \ref{fig:NewFEA}, we calculate T$_1$ of cell L02 with different depolarization probabilities, which are carefully chosen to correspond to realistic surface conditions: depolarization probability of the O-ring valve is much higher than that of the pyrex glass surface; the spherical volume has the lowest depolarization probability because of its rubidium coating.

Fig. \ref{fig:NewFEA}(a) shows that depolarization on the valve can be neglected as designed; it is consistent with the preceding estimation of (T$_1$)$_{val}$ = 26 hours at 0.3 atm for the new cells. It also shows that depolarization on the capillary tube surface is pressure dependent for any value of pressure, which not only is contrary to the experimental data in the high pressure region, but also can not explain the sharp decrease of the T$_1$ observed at the pressure below 0.3 atm. Relaxation due to the spherical volume is considered in the calculations shown in \ref{fig:NewFEA}(b), the calculated T$_1$ curve is flattened, but it deviates from the experimental data in both the high pressure region and the low pressure region. Our FEA study shows that the pressure dependent relaxation observed in cell L02 can not be solely ascribed to depolarization on any surface together with the diffusion, indicating other possible sources of relaxation giving rise to the observed pressure dependence.

Other possible candidates proposed in \cite{Saam}, such as an inhomogeneous ac field, to explain the linear pressure dependence at low temperature, can be safely ruled out. To reach the relaxation rate measured in the experiment, the magnitude of this ac field needs to be such that the SQUID sensor can not survive in this environment \cite{Cates}. This is not supported by results at low temperature using SQUID reported in \cite{Zheng2}. Sealed cells can yield very long T$_1$ at either low pressures or high pressures. However, it is difficult to quantitatively compare one sealed cell with another at different pressures, since different cells have different surface conditions. The re-fillable cells used in our study, provide an excellent way to evaluate the relationship between the wall relaxation and pressure, and similar T$_1$ has always been observed no matter how many times the cell is re-filled under normal conditions. The linear pressure dependence has also been consistently observed in our measurements with different geometries and experimental conditions.

Similar pressure dependence was observed in references \cite{Chapman,Lusher,Chen} as well. However, they either ascribe the behavior to the pressure dependence of the depolarization probability \cite{Chapman} or just presented the result without an explanation \cite{Lusher,Chen}. To explain the pressure dependent T$_1$ observed in \cite{Zheng2}, Saam \textit{et al.} \cite{Saam} proposed that the diffusion associated with the relaxation on the valve could contribute dominantly to T$_1$ and result in a pressure dependent behavior. Using the boundary condition given by \cite{Saam}, we have carried out FEA studies of the new measurements as well as the room temperature measurement reported in \cite{Zheng2}. The FEA results are not in agreement with the room temperature data in terms of the overall behavior of the T$_1$-pressure relationship. In the new measurements at room temperature, we use cells with significantly lengthened capillary tube to suppress the relaxation due to depolarization on the valve surface. The pressure dependent T$_1$ is still observed with these new cells, indicating that the depolarization on the valve can not explain this behavior, nor can the depolarization from the long capillary tube. All evidence points to the fact that a gap exists between theory and experimental observation and the observed pressure dependent T$_1$ remains an open question. Other unknown sources of pressure dependent relaxation are at play, and
 need to be identified in order to understand the results. This linear pressure dependence is interesting and nontrivial, and is worthy of further investigation.
\\

Special thanks to J. Rishel for making all the cells reported in this paper, and T. Gentile, S. Jawalkar, G. Laskaris and X. Yan for helpful discussion. This work is supported by the U.S. Department of Energy under contract number DE-FG02-03ER41231 and Duke University.


\begin{thebibliography}{99}
\bibitem{Zheng2}W. Zheng, H. Gao, Q. Ye, and Y. Zhang, \Journal{\PRA}{83}{061401(R)}{2011}.
\bibitem{Saam}B. Saam, A. K. Petukhov, J. Chastagnier, T. R. Gentile, R. Golub, and C.M. Swank, \Journal{\PRA}{85}{047401}{2012}.
\bibitem{Xu}W. Xu \textit{et al.}, \Journal{\PRL}{85}{2900}{2000}.
\bibitem{Middleton2}H. Middleton \textit{et al.}, \Journal{\MRM}{33}{271}{1995}.
\bibitem{Petukhov}A.K. Petukhov, G. Pignol, D. Jullien, and K.H. Andersen, \Journal{\PRL}{105}{170401}{2010}.
\bibitem{Zheng}W. Zheng, H. Gao, B. Lalremruata, Y. Zhang, G. Laskaris, W. M. Snow, and C.B. Fu,\Journal{\PRD}{85}{031505(R)}{2012}.
\bibitem{Chu}P.-H. Chu, A. Dennis, C.B. Fu, H. Gao, R. Khatiwada, G. Laskaris, K. Li, E. Smith, W. M. Snow, H. Yan, and W. Zheng, \Journal{\PRD}{87}{011105(R)}{2013}.
\bibitem{Newbury}N.R. Newbury \textit{et al.},
\Journal{\PRA}{48}{4411}{1993}.
\bibitem{Mullin}W.J. Mullin, F. Laloe\:, and M.G. Richards, \Journal{\LTP}{80}{1}{1990}.
\bibitem{Cates}G.D. Cates \textit{et al.},
\Journal{\PRA}{37}{2877}{1988}.
\bibitem{Garwin}R.L. Garwin, and H.A. Reich, \Journal{\PR}{115}{1478}{1959}.
\bibitem{Jacob}R.E. Jacob, S.W. Morgan, and B. Saam, \Journal{\JAP}{92}{1588}{2002}.
\bibitem{Chapman}R. Chapman, and M.G. Richards, \Journal{\PRL}{33}{18}{1974}.
\bibitem{Lusher}C.P. Lusher, M.F. Secca, and M.G. Richards, \Journal{\LTP}{72}{25}{1988}.
\bibitem{Chen}H.H. Chen \textit{et al.}, \Journal{\PRA}{81}{033422}{2010}.

\end{thebibliography}
\end{document}